\documentclass[final,5p,times,twocolumn]{elsarticle}

\usepackage{amssymb}
\usepackage{amsmath}
\usepackage{multirow}
\usepackage{xcolor}
\usepackage{soul}
\usepackage{upgreek}

\begin{document}

\begin{frontmatter}

\title{Constraints on dark photon dark matter using Voyager magnetometric survey}

\author[LPSC]{G.~Pignol}         \ead{pignol@lpsc.in2p3.fr}
\author[LPSC]{B.~Clement}
\author[LPSC]{M.~Guigue}
\author[LPSC]{D.~Rebreyend}
\author[LPSC]{B.~Voirin}

\address[LPSC]{LPSC, Universit\'e  Grenoble Alpes, CNRS/IN2P3, Grenoble, France}

\begin{abstract}
The dark photon, an new hypothetical light spin 1 field, constitutes a well-motivated dark matter candidate. It manifests as an oscillating electric field with a fixed direction, which can be observed in magnetometric records. In this letter, we use magnetometer data from the Voyager probes to look for the dark photon in the $10^{-24}$ eV to $10^{-19}$ eV mass range, corresponding to frequencies between $10^{-9}$ Hz and $10^{-4}$ Hz. We also discuss the sensitivity of possible future SQUID magnetometry experiments.
\end{abstract}

\begin{keyword}

Dark matter \sep dark photon \sep Voyager probes \sep interplanetary magnetic fields


\end{keyword}

\end{frontmatter}


\section{Introduction}
\label{Introduction}

We now have a compelling set of evidence for the existence of some form of cold, non-baryonic dark matter 
exerting gravitational influence from astronomical to cosmological scales. 
Elucidating the microscopic nature of the dark matter 
is one of the most pressing challenges in physics. 
This important open question connects particle physics and cosmology. 
It is likely that future progress in one field will proceed from the advance in the other. 
Among the countless theories to explain the dark matter, 
two broad classes can be distinguished: WIMPs and WISPs.

WIMPs are weakly interacting massive particles, with a typical mass in the GeV to TeV range. 
They are still actively searched for by direct detection with low background detectors burried underground. 
Alternatively, the particle could be produced at the LHC in high energy proton-proton collisions, 
it is looked for in events with large missing energy. 

WISPs are weakly interacting slim particles whose mass can be much less than 1 eV 
(see \cite{Arias2012,Ringwald2012} for recent reviews). 
This type of dark matter is necessarily bosonic, 
since the Pauli exclusion principle forbids fermions to condense above a certain density, the so-called Tremaine-Gunn bound \cite{Tremaine1979}. 
Possibly, these bosonic fields are produced in the early Universe via the vacuum misalignment mechanism \cite{Preskill1983,Abbott1983}. 
They behave very much like classical fields, oscillating at a frequency $f = M c^2 / \hbar$, 
where $M$ is the mass of the bosonic field. 
Although the particles are very light as compared to the temperature of the early Universe, the field behaves as cold dark matter. 
More precisely, the equation of state of the energy density stored in the oscillations is the same as that of pressureless dust. 

The scenario of dark matter as a condensate of very light bosons has mostly been discussed in the framework of the QCD axion, a spin 0 field invented to solve the strong CP problem \cite{Preskill1983,Abbott1983}. 
Theory predicts that dark matter axions could be converted into microwave photons in the presence of an external magnetic field. 
Haloscopes, i.e. devices realizing this conversion such as ADMX \cite{Asztalos2010}, are probing a large portion of plausible axion masses and couplings. 

Recently, another form of WISPy dark matter has been proposed by Nelson and Scholtz \cite{Nelson2011}. 
The scenario consists in adding a new vector field associated with a hidden $U(1)$ gauge symmetry called the dark photon. 
The energy stored in the oscillations of this field would constitute the dark matter. 
In this letter we explore the possibility of detecting these oscillations using precise magnetometers. 
The proposed technique could be applied in a broad frequency domain ranging from $10^{-9}$~Hz to $1$~kHz, corresponding to dark photon masses in the range $10^{-24}$ - $10^{-12}$~eV. 
It is thus complementary to the newly proposed search with a dish antenna in the radio domain \cite{Horns2012}. 
The paper is organised as follows. 
In section \ref{dark photon} we review the dark photon dark matter scenario and derive the expected magnetic signal. 
In section \ref{Voyager} we set a bound on the kinetic mixing $\chi$ using the magnetometric data recorded by the two Voyager interplanetary missions. 
Then, the sensitivity of a dedicated measurement with SQUID magnetometers is discussed in section \ref{Conclusion}.

\section{Magnetic signature of dark photon dark matter}
\label{dark photon}

We consider an extension of standard electrodynamics with an additional spin 1 field $X_\mu$ \cite{Okun1982}, defined by the Lagrangian density
\begin{eqnarray}
\mathcal{L} & & = -\frac{1}{4} F_{\mu \nu} F^{\mu \nu} - \frac{1}{4} X_{\mu \nu} X^{\mu \nu} + \frac{M^2}{2} X_\mu X^\mu \\ 
\nonumber
 & & + \frac{\chi}{2} F_{\mu \nu} X^{\mu \nu} + J^\mu A_\mu
\end{eqnarray}
where $F_{\mu \nu} = \partial_\mu A_\nu - \partial_\nu A_\mu$ represents the field strength of the ordinary photon field $A_\mu$, 
$J^\mu$ is the usual current of charged fermions, 
and $X_{\mu \nu} = \partial_\mu X_\nu - \partial_\nu X_\mu$ is the field strength of the new boson of mass $M$. 
The name of the new boson is not unique in the litterature, it was first referred to as a \emph{paraphoton}, then more recently as \emph{hidden photon} or \emph{dark photon}. 
Since we are regarding the new field as a dark matter candidate we choose to use the latter name. 

The dark photon couples to the rest of the world through the kinetic mixing term with the dimensionless coupling $\chi$. 
String-inspired extensions of the Standard Model of particle physics typically predict a value for $\chi$ in the range $10^{-12} - 10^{-3}$ \cite{Jaeckel2010}. 
By making the transformation $A_\mu = \tilde{A}_\mu + \chi X_\mu$ it is apparent that a particle of charge $e$ for the usual photon has a suppressed effective charge of $\chi e$ for the dark photon. 

The classical equations of motion for the dark photon field
admit oscillating solutions of the form
\begin{equation}
X_i = Re \left( x_i e^{- i M t} \right) \quad , \quad X_0 = 0, 
\end{equation}
where $x_i$ are complex amplitudes. 
The corresponding complex amplitude for the dark electric field $ - \partial_0 \vec{X}$ is $\vec{E}_{\rm dark} = i M \vec{x}$. 
The energy density stored in the oscillation is $\rho = \frac{M^2}{2} |\vec{x}|^2$. 
In terms of the electric field, in SI units we have
\begin{equation}
\rho = \frac{\epsilon_0}{2} |\vec{E}_{\rm dark}|^2.  
\end{equation}

Following Nelson and Scholtz \cite{Nelson2011} we attribute the energy density of dark matter to the oscillations of the dark photon. 
We use the canonical value of $\rho = 0.3~{\rm GeV/cm}^3$ \cite{PDG2014} for the local dark matter density and find
$E_{\rm dark} = 3300~{\rm V/m}$. 

This dark electric field oscillates in the rest frame of the dark matter. 
Our motion relative to the halo induces a motional dark magnetic field with complex amplitude given by 
\begin{equation}
\vec{B}_{\rm dark} = - \frac{\vec{v}}{c^2}\times \vec{E}_{\rm dark}, 
\end{equation}
where $\vec{v}$ is the relative velocity of the Earth in the halo, 
that we assume to point to the Cygnus constellation. 
We use the canonical value for solar velocity $v = 254~{\rm km/s}$. 

The dark magnetic field couples to the usual electric charges with the coupling constant scaled by $e \rightarrow \chi e$. 
In other terms, we should search for a magnetic field $\vec{B} = \chi \vec{B}_{\rm dark}$, 
oscillating at the frequency $f = Mc^2/\hbar$, 
in a plane transverse to $\vec{v}$, with an amplitude of 
\begin{equation}
B = \chi \sin(\theta) v E_{\rm dark}/c^2 = 9.3 \ \chi \sin(\theta)~{\rm nT}, 
\end{equation}
where $\theta$ is the angle between our velocity $\vec{v}$ and the direction of the dark electric field.


\section{Constraint from the Voyager magnetic probe}
\label{Voyager}

In order to search for the signal described above in the low-frequency regime (about $10^{-9}$ Hz or a mass of around $10^{-22}$ eV), we looked for long-running magnetic experiments in low noise environments. One such experiment is the Voyager Program.
The two Voyager probes were constructed by the Jet Propulsion Laboratory and launched in 1977. Among many other instruments, they carried triaxial fluxgate magnetometers \cite{Behannon1977}. The last planet encountered by Voyager 1 was Saturn in 1980, whereas Voyager 2's last flyby was above Neptune in 1989. Between these events and their arrival at the termination shock, respectively in 2004 for Voyager 1 and in 2007 for Voyager 2, the probes were deep in interplanetary space. 

\begin{figure} 
\centering
\includegraphics[width=\linewidth]{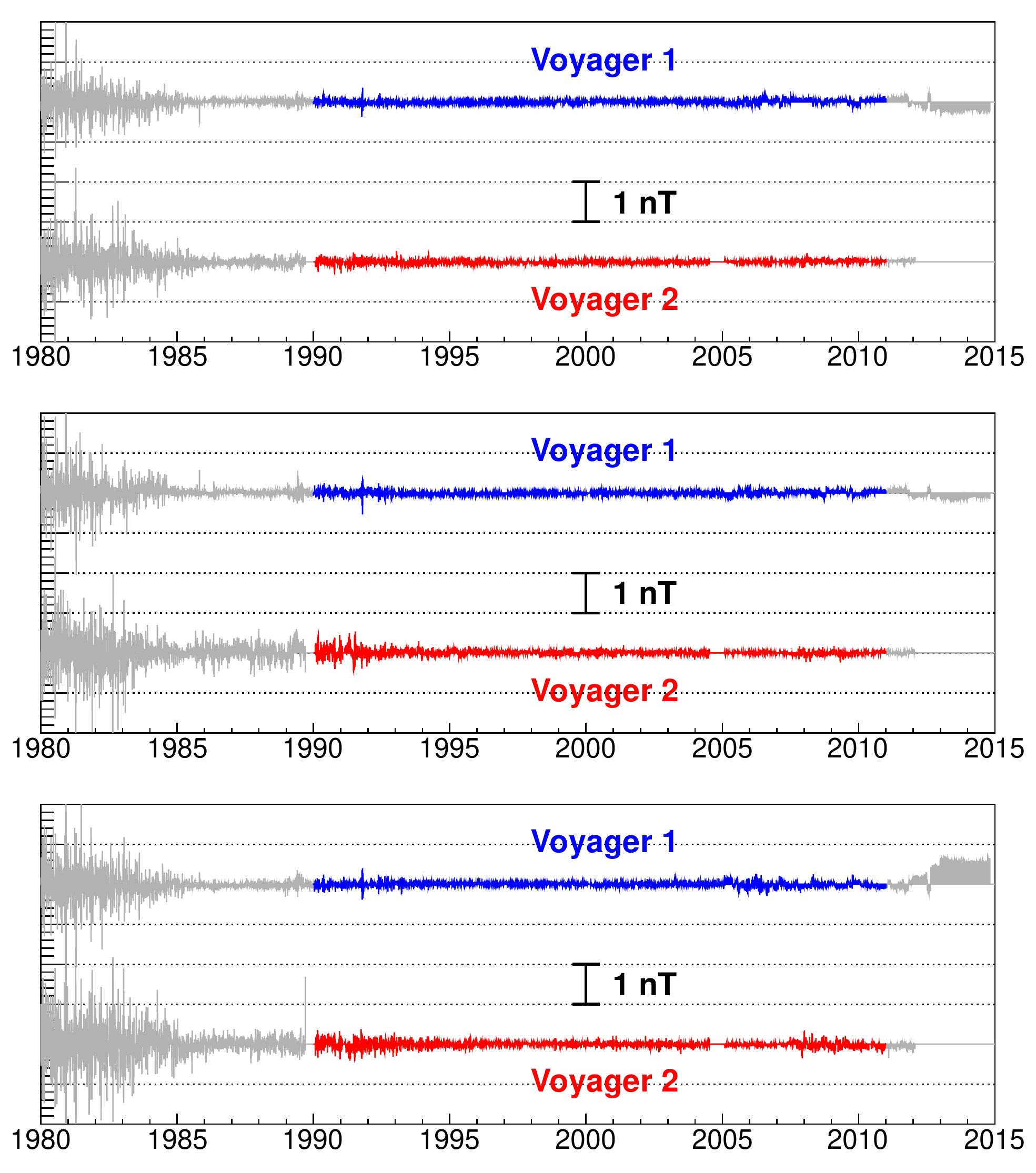} 
\caption{
Magnetic field values as a function of time. The colored parts are the the data actually used, running from 1990 to 2011. The blue line is the data from Voyager 1, the red one from Voyager 2. The upper figure show the magnetic field along the $x$ (apical) axis, the middle figure along the $y$ axis and the bottom figure along the $z$ axis. For readability, only one in ten points is actually shown.
}
\label{showTime}
\end{figure}
\begin{figure}
\centering
\includegraphics[width=\linewidth]{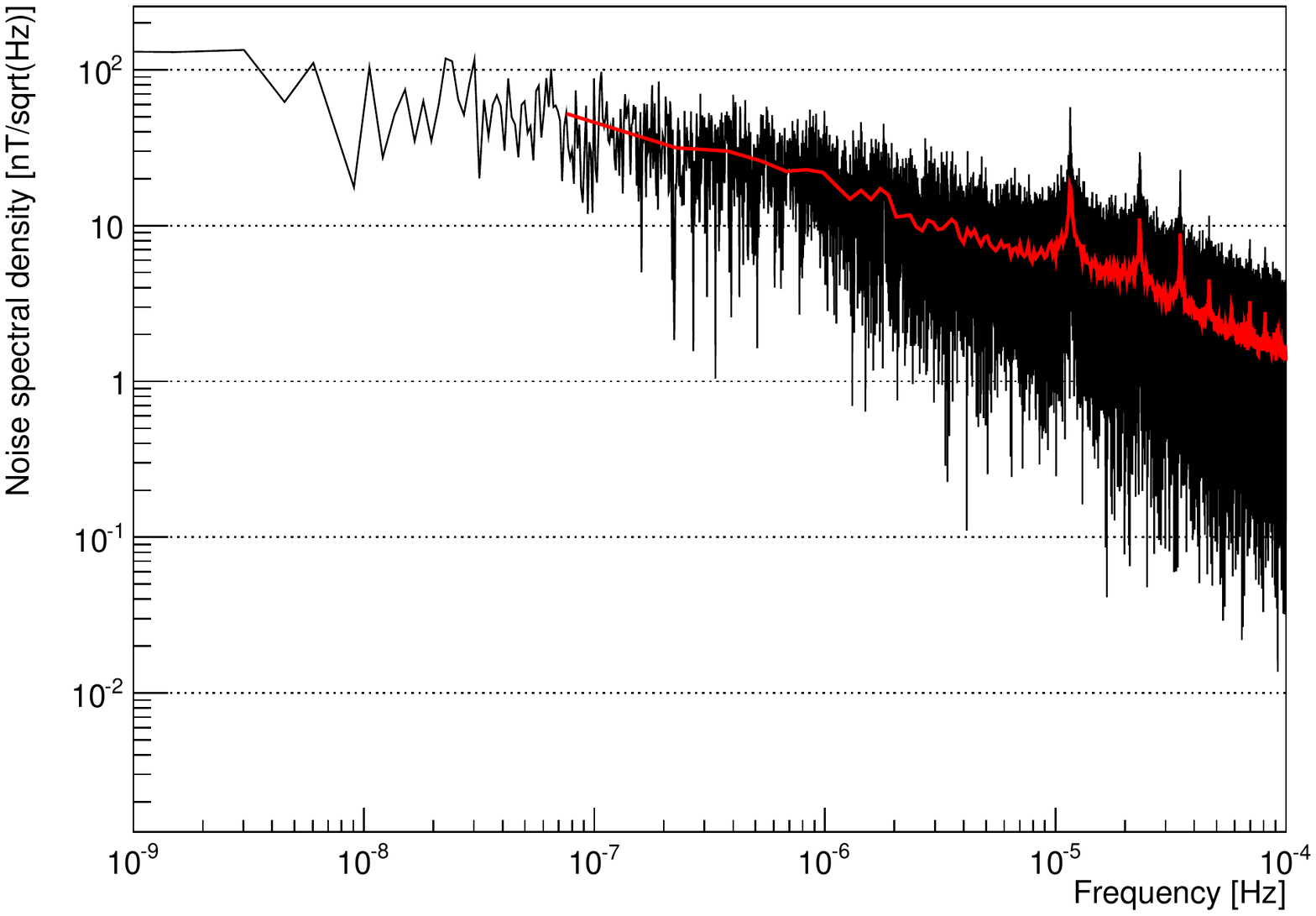}
\caption{
Noise spectral density of the transverse magnetic field (mean data of the two probes). The black line is the signal proper, the red line is a moving average calculated on 100 points. 
}
\label{fftVy1}
\end{figure}

 
The data set running from 1990 to 2011 presented a noise below 1 nT and was very well suited to our needs. The magnetic field seen by both probes during their entire flight is presented in figure \ref{showTime}. From now on, we will exclusively use magnetic data taken between 1990 and 2011. The data were taken from the Goddard Space Flight Center's OmniWeb and FTP services. It consists of triaxial hourly samples of the external magnetic field. These values are given in RTN coordinates (Radial Tangential Normal); it is then necessary to transform them into a referential more adapted to our aims. There is clearly a preferred direction which is along the speed of the probes in the galactic referential; this direction is essentially that of the speed of the Sun in the galaxy, also known as the {\it solar apex}. Along this axis, there can be no signal from the dark photon, and thus the corresponding magnetometric data component can be used to discriminate oscillations resulting from sources other than the dark photon.
This direction is called the {\it apical direction} or $x$ axis, the plane orthogonal to it is called the {\it transverse} plane. We have then selected two other axes in order to get a direct trihedron; those are arbitrary in the sense that our results will not depend on this choice. We call them the $y$ and $z$ axes. For our purposes, we have chosen a $y$ axis in the galactic plane, pointing in the direction of positive galactic longitude and orthogonal to the solar apex, and a $z$ axis completing the orthonormal coordinate system.
In order to use the data of both probes, we will work with the mean of the magnetic values of Voyager 1 and Voyager 2 for each hour and each axis. The magnetic values in the tranverse plane are coded into a complex number of form $s(n) = \frac{B_{1y}+B_{2y}}{2}+i\frac{B_{1z}+B_{2z}}{2}$, where $B_{1y}$ is the component of the field along the $y$ axis seen by Voyager 1 at sample time number n, and other values are labeled accordingly. We then compute the discrete Fourier transform of this signal, given by :
\begin{equation}
S(k) = \sum_{0}^{N-1} s(n) \: e^{-2ik\pi \frac{n}{N}},
\end{equation}
where $S$ is the fourier transform, $s$ is the signal described above, and $N$ is the number of points considered in the transform.
Once we have this result, it is easy to get the noise spectral density of the signal, given by $S_{x}(k) = \sqrt{ \frac{ \lvert S(k) \times \Delta t \rvert^{2}}{T}}$, with $\Delta t$ the sampling period and $T$ the total duration of the acquisition. Since the noise density depends only on the module of the Fourier transform, it is independant of the explicit choice of the $y$ and $z$ axes. The results are given in the figure \ref{fftVy1}.

\begin{figure}
\centering
\includegraphics[width=\linewidth]{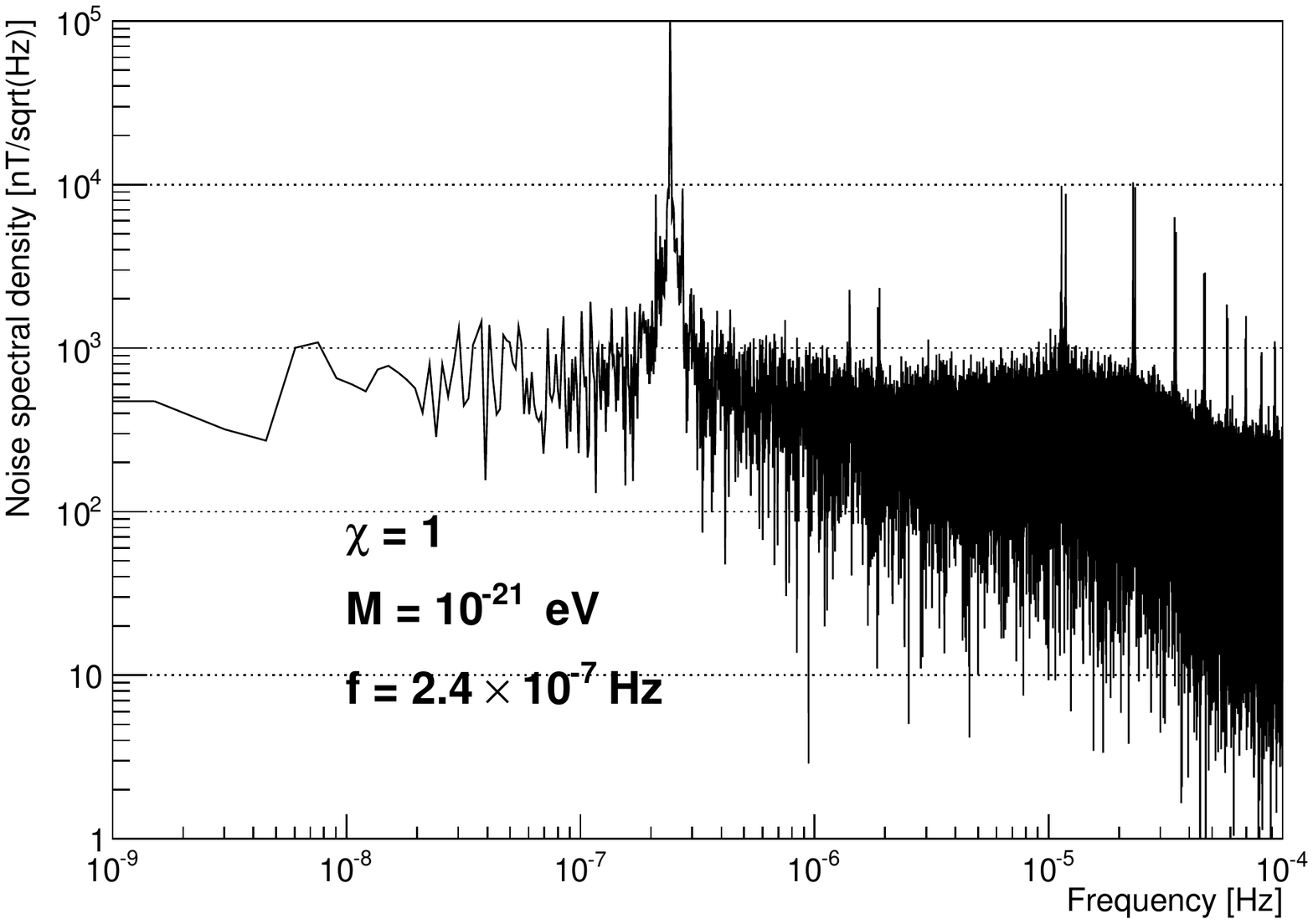}
\caption{
Noise spectral density of a simulated dark photon signal with parameters M and $\chi$ and distorsion induced by the missing data. 
}
\label{missingData}
\end{figure}

It is immediately apparent that there is a one day period ($1.1 \times 10^{-5}$ Hz) signal in the data, along with associated harmonics. This is by no means a sign of the dark photon, as the same periodicity appears in the apical field. This peculiar oscillation stems from the values missing in the record. The figure \ref{missingData} presents a simulated case with a $10^{-21}$ eV and $\chi = 1$ dark photon, with distorsion induced by the points missing on the record. The expected mass peak at $2.4 \times 10^{-7}$ Hz is present, but we see  24-hour components and associated harmonics arising as well. Those frequencies remain even if we get rid of any signal and simply fill any non-missing data with a constant value. Thus, we can safely assume that those peaks are just an artifact. We can observe that other than these previously mentioned frequencies, no significant peak can be seen.

\begin{figure}
\centering
\includegraphics[width=\linewidth]{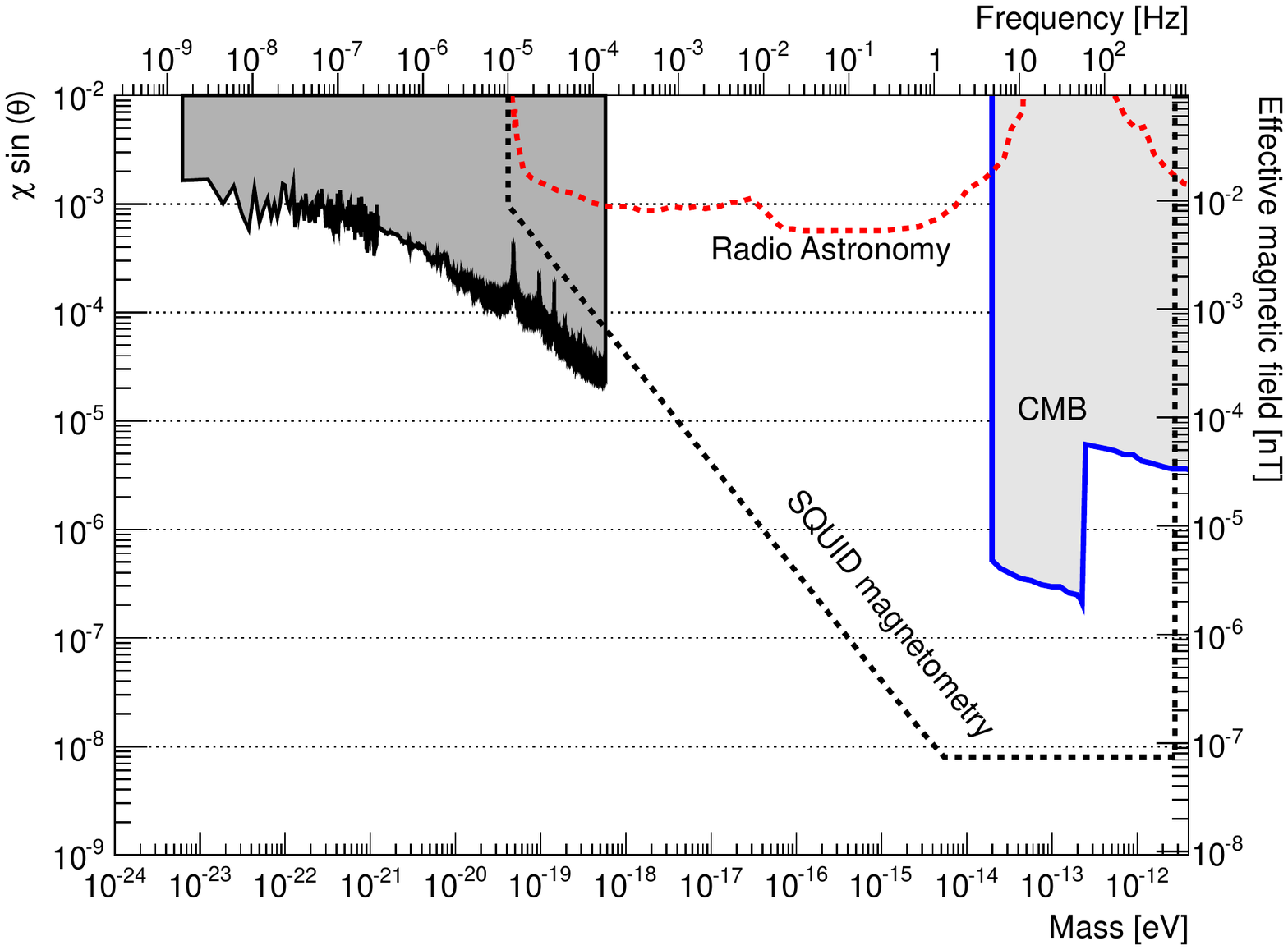}
\caption{
Presentation of exclusion limits on $\chi$ as a function of the mass of the dark photon. Black solid line : this work. Red dotted line : expected constraints from a combination of radio observations \cite{Lobanov2013}. Blue solid line : Constraints from Cosmic Microwave Background observations \cite{Mirizzi2009}. Black dotted line : SQUID magnetometry expected limit, explained in section \ref{Conclusion}.
}
\label{limite}
\end{figure}

 The next step is then to determine an exclusion limit of the values of $\chi \sin(\theta)$. Since the noise level varies with the frequency, no single criterion of exclusion can be used on the integrality of the bandwidth. Instead, we compute a moving average of the noise spectral density by blocks of 200 points (corresponding to a bandwidth of $3.0 \times 10^{-7}$ Hz), along with the standard deviation of these blocks. Having verified that the noise is gaussian in nature, with 95 \% confidence, we exclude any signal with a given frequency and an amplitude larger than the mean at this frequency plus two standard deviations. For low frequencies, this method is inadapted, as the noise levels vary greatly from one point to the next. For the 200 first points (from $1.5 \times 10^{-9}$ Hz to $3.0 \times 10^{-7}$ Hz), we exclude any signal greater than the noise spectral density of the data at a given frequency plus two times the standard deviation of these 200 first points.  In order to do that, we simulate a dark photon oscillating signal with an arbitrary frequency and a value of the coupling parameter $\chi$ set to 1, and we compute the noise spectral density, which has a maximum value of $2.12 ~ \mathrm{nT}/ \! \sqrt{\mathrm{Hz}}$. This density does not depend on the mass of the particle to first order, and is linear in $\chi$, which allows us to transform the previous limit on the noise density into a limit on $\chi \sin(\theta)$. The final limit is represented on figure \ref{limite}, compared to expected limits from other experiments. For completeness, the same analysis has been extended to the apical field; results are given in table \ref{effectiveFields} in the form of constraints on the effective field amplitudes.

\begin{table}
\center
\caption{
Constraints (at 95 \% C. L.) on the amplitude of an oscillating magnetic field in apical and transverse directions 
in the frequency range accessible with Voyager magnetic data. 
}
\begin{tabular}{l l l}
\hline
Frequency	& $B_{\rm T}$ & $B_{\rm A}$ \\
\hline 
$10^{-8}$ Hz & 16 pT & 9   pT \\
$10^{-7}$ Hz & 11 pT & 7 pT \\
$10^{-6}$ Hz & 4 pT & 2.2 pT \\
$10^{-5}$ Hz & 1.5 pT & 0.7 pT \\
$10^{-4}$ Hz & 0.3 pT & 0.2 pT \\
\hline 
\end{tabular}
\label{effectiveFields}
\end{table}

\section{Conclusion and prospects}
\label{Conclusion}

We have shown that precision magnetometry could be used to search for the dark photon, a newly proposed form of dark matter. 
In this scenario, the dark matter is attributed to the coherent oscillation of the new field, which acts as a pseudo magnetic field due to our motion in the halo. 
The very long magnetic records from the Voyager 1 and 2 space probes were used to set competitive limits 
in the mass range $10^{-23}$ eV to $10^{-18}$ eV, corresponding to frequencies between about $10^{-9}$ Hz and $10^{-4}$ Hz. Interestingly, this covers the scenario known as Fuzzy dark matter \cite{Hu2000}. With 95 \% confidence, we can exclude values of the coupling parameter $\chi$ as low as $3 \times 10^{-5}$ for the highest masses, which represents the best limit for this range of frequencies. 

In the future, a natural development would be to look at somewhat shorter periods. It is sensible to propose exploration of such periods (from the year to the millisecond) with a similar approach, using SQUID (Superconducting Quantum Interference Device) magnetometry. Using noise data published in Henry {\it et al.} \cite{Henry2008}, which gives a noise spectral density of $4.0 ~\rm{fT}/ \! \sqrt{\rm{Hz}}$ above 1 Hz and of $(4.9/f) ~\rm{fT}/ \! \sqrt{\rm{Hz}}$, we suppose a data acquisition for a time of one month and that we can see a signal if its noise spectral density is two times as large as that of the ambient noise. We can then expect a sensitivity as low as $\chi = 10^{-8}$. The predicted limit is entitled "SQUID magnetometry" in figure \ref{limite}. More generally, any precision magnetometry experiment, for example a clock comparison experiment \cite{Altarev2009}, could be adapted to search for these oscillations. These prospects open new perspectives in the hunt for the dark matter.

\section*{Acknowledgments}

We thank Ann Nelson for useful discussions. 
This work was supported by "Investissements d'avenir, Labex ENIGMASS". 
We acknowledge use of NASA/GSFC’s Space Physics Data Facility’s ftp service, and OMNI data.


\begin{thebibliography}{00}

\bibitem{Arias2012}
P.~Arias {\it et al}, 
JCAP {\bf 1206} (2012) 013. 

\bibitem{Ringwald2012}
A.~Ringwald, 
Phys. Dark Univ.  {\bf 1} (2012) 116. 

\bibitem{Tremaine1979}
S.~Tremaine and J.~E.~Gunn, 
Phys. Rev. Lett. {\bf 42} (1979) 407. 

\bibitem{Preskill1983}
J.~Preskill, M.~B.~Wise and F. Wilczek, 
Phys. Lett. B {\bf 120} (1983) 127. 

\bibitem{Abbott1983}
L.~F.~Abbott and P.~Sikivie, 
Phys. Lett. B {\bf 120} (1983) 133. 

\bibitem{Asztalos2010}
S.~J.~Asztalos {\it et al}, 
Phys. Rev. Lett. {\bf 104} (2010) 041301. 

\bibitem{Nelson2011}
A.~E.~Nelson and J.~Scholtz, 
Phys. Rev. D {\bf 84} (2011) 103501.

\bibitem{Horns2012}
D.~Horns {\it et al}, 
JCAP {\bf 1304} (2013) 016. 

\bibitem{Okun1982}
L.~B.~Okun, 
Sov. Phys. JETP {\bf 56} (1982) 502. 

\bibitem{Jaeckel2010}
J.~~Jaeckel and A.~Ringwald, 
Ann. Rev. Nucl. Part. Sci. {\bf 60} (2010) 405.

\bibitem{PDG2014}
K.A.~Olive {\it et al} (Particle Data Group), 
Chin. Phys. C, {\bf 38}, (2014) 090001. 

\bibitem{Behannon1977}
K.W.~Behannon {\it et al},
Space Science Reviews {\bf 21} (1977) 235.

\bibitem{Lobanov2013}
A.~P.~Lobanov, H-S. Zechlin and D. Horns,
Phys. Rev. D {\bf 87} (2013) 065004.

\bibitem{Mirizzi2009}
A.~Mirizzi, J.~Redondo and G.~Sigl,
JCAP {\bf 03} (2009) 026.

\bibitem{Hu2000}
W.~Hu, R.~Barkana and A.~Gruzinov,
Phys. Rev. Letters {\bf 85} (2000) 1158.

\bibitem{Henry2008}
S.~Henry {\it et al.},
JINSI {\bf 3} (2008) P11003.

\bibitem{Altarev2009}
I.~Altarev {\it et al.},
Phys. Rev. Letters {\bf 103} (2009) 081602.



\end{thebibliography}
\end{document}